\documentclass[12pt]{article} 
\usepackage{amsfonts}
\usepackage{amsmath}
\usepackage{latexsym}
\usepackage{amssymb}
\usepackage{txfonts}
\usepackage[american]{babel}
\makeatletter
\def\timenow{%
\@tempcnta=\time \divide\@tempcnta by 60 \number\@tempcnta:\multiply
\@tempcnta by 60 \@tempcntb=\time \advance\@tempcntb by -\@tempcnta
\ifnum\@tempcntb <10 0\number\@tempcntb\else\number\@tempcntb\fi}

\newcounter{outputpage}

\renewcommand{\@oddhead}
{\stepcounter{outputpage}\hfill\hfill\theoutputpage}

\renewcommand{\@evenhead}
{\stepcounter{outputpage}\hfill\hfill\theoutputpage}

\renewcommand{\@oddfoot}
{\vbox{
\hrule
\vspace{3pt}
\hfil
{\scriptsize\textit{
\hfill\hfill\jobname.tex; \today; \timenow; p. \theoutputpage}}
\hfil
}}

\renewcommand{\@evenfoot}
{\vbox{
\hrule
\vspace{3pt}
\hfil
{\scriptsize\textit{
\hfill\hfill\jobname.tex; \today; \timenow; p. \theoutputpage
}}
\hfil
}}

\makeatother

\def\RR{{\mathbb R}} 
\def\CC{{\mathbb C}} 
 
\def\ZZ{{\mathbb Z}}

\def\cV{{\cal V}}

\def\tr{\mathrm{ tr\,}} 
\def\Tr{\mathrm{ Tr\,}}

\def\Real{\mathrm{Re\,}}
\def\Imag{\mathrm{Im\,}}

\def\dim{\mathrm{dim\,}}
 
\def\vol{\mathrm{ vol\,}}

\def\End{\mathrm{End}}

\def\be{\begin{equation}} 
\def\ee{\end{equation}} 
\def\bea{\begin{eqnarray}} 
\def\eea{\end{eqnarray}} 
\def\bed{\begin{definition}{\ }}
\def\eed{\end{definition}}
\def\bd{\begin{description}}
\def\ed{\end{description}}
\def\bc{\begin{center}}
\def\ec{\end{center}}

\newtheorem{lemma}{Lemma}

\newtheorem{definition}{Definition}

\def\sideremark#1{\ifvmode\leavevmode\fi\vadjust{\vbox to0pt{\vss
\hbox to 0pt{\hskip\hsize\hskip1em
\vbox{\hsize2cm\tiny\raggedright\pretolerance10000
\noindent #1\hfill}\hss}\vbox to8pt{\vfil}\vss}}}

\begin{document}

\begin{titlepage}

\thispagestyle{empty}
\null
%\vspace{-10mm}
%\hspace*{50truemm}{\hrulefill}\par\vskip-4truemm\par
%\hspace*{50truemm}{\hrulefill}\par\vskip5mm\par
%\hspace*{50truemm}{{\large\sc New Mexico Tech {\rm 
%(\cdate)}}}\vskip4mm\par
%\hspace*{50truemm}{\hrulefill}\par\vskip-4truemm\par
%\hspace*{50truemm}{\hrulefill}
%\par
%\bigskip
%\bigskip
%\par
%\rightline{\LARGE\textbf{\textsf{DRAFT}}}
\par
\vspace{3cm}
\centerline{\huge\bf Quantum Heat Traces}
%\bigskip
%\bigskip
%\centerline{\huge\bf and Bogolyubov Invariant}
\bigskip
\bigskip
\centerline{\Large\bf Ivan Avramidi{
%\footnote{Alexander von Humboldt Fellow}
%\footnote{E-mail: ivan.avramidi@nmt.edu}
\footnote{On leave of absence from New Mexico Institute of Mining and Technology,
Socorro, NM 87801, USA}}
}
\bigskip
%\centerline{\it Department of Mathematics}
\centerline{\it Mathematical Institute, University of Bonn}
\centerline{\it Endenicher Allee 60, D-53115, Bonn, Germany}
\centerline{\it E-mail: ivan.avramidi@nmt.edu}
\bigskip
%\centerline{Revised on }
\medskip
%\maketitle 
\vfill

{\narrower
\par
We study new invariants of elliptic partial differential operators 
acting on sections of a vector bundle over a closed Riemannian manifold
that we call the relativistic heat trace and the quantum heat traces.
We obtain some reduction formulas expressing these new invariants in terms
of some integral transforms of the usual classical heat trace and
compute the asymptotics of these invariants. The coefficients of these
asymptotic expansion are determined by the usual heat trace coefficients
(which are locally computable) as well as by some new global invariants.

\par}

\vfill

\end{titlepage}

%=============================================================
%=================================================================
 
\section{Introduction}
\setcounter{equation}0

%=================================================================

The heat kernel is one of the most important tools of global analysis, spectral
geometry, differential geometry and mathematical physics
\cite{hadamard23,berline92,gilkey95,hurt83}, in particular,
quantum field theory, even
financial mathematics \cite{avramidi15}. In quantum field
theory the main objects of interest are described by the Green functions of
self-adjoint elliptic partial differential operators on manifolds and their
spectral invariants such as the functional determinants. In spectral geometry
one is interested in the relation of the spectrum of natural elliptic partial
differential operators to the geometry of the manifold.
There are also non-trivial links between the spectral invariants and the
non-linear completely integrable evolution systems, such as Korteweg-de Vries
hierarchy (see, e.g. 
\cite{hurt83,avramidi95b,avramidi00,avramidi00b}). 
In many interesting cases such
systems are, in fact, infinite-dimensional Hamiltonian systems, and the
spectral invariants of a linear elliptic partial differential operator are
nothing but the integrals of motion of this system.
In financial mathematics the behavior of the derivative securities
(options) is determined by some deterministic parabolic partial differential
equations of diffusion type with an elliptic partial differential operator of
second order. The
conditional probability density is then nothing but the fundamental solution of
this equation, in other words, the heat kernel  
\cite{avramidi15}.

Instead of studying the spectrum of a differential operator directly one
usually studies its spectral functions, that is, spectral traces of some
functions of the operator, such as the zeta function, and the heat trace.
Usually one does not know the spectrum exactly; that is why, it becomes very
important to study various asymptotic regimes. It is well known, for example,
that one can get information about the asymptotic properties of the spectrum by
studying the short time asymptotic expansion of the heat trace. The
coefficients of this expansion, called the heat trace coefficients (or global
heat kernel coefficients), play very important role in spectral geometry and
mathematical physics \cite{hurt83,gilkey95}.

The existence of non-isometric isospectral manifolds
demonstrates that the spectrum alone does not determine the geometry
(see, e.g. \cite{berger03}). That is why, it makes sense 
to study more general invariants of
partial differential operators, maybe even such invariants that are
not spectral invariants, that is, invariants that 
depend not only on the eigenvalues but also on the eigenfunctions, and,
therefore, contain much more information about the geometry of the manifold.

The case of a Laplace operator on a compact manifold without boundary
is well understood and there is a vast literature on this subject, see
\cite{gilkey95} and the references therein.
In this case there is a
well defined local asymptotic expansion of the heat kernel, which enables one
to compute its diagonal and then the heat trace by directly integrating the
heat kernel diagonal; this gives all heat trace coefficients. 
The heat trace asymptotics of laplace type operators
have been extensively
studied in the literature, and many important results have been discovered. The
early developments are summarized in the books \cite{gilkey95,berline92} with
extensive bibliography, see also \cite{avramidi00,avramidi99,avramidi02,avramidi10}. 

We initiate the study of new invariants of second-order
elliptic partial
differential operators acting on sections of vector bundles over 
compact Riemannian
manifolds without boundary. The long term goal of this
project is to develop a comprehensive methodology for such invariants in the
same way as the theory of the standard heat trace invariants.  
We draw a deep analogy
between the spectral invariants of elliptic operators and the classical and
quantum statistical physics \cite{landau69}. 
We consider am elliptic self-adjoint positive partial differential operator
$H$ and its square root,
$\omega=H^{1/2}$, which is an elliptic
self-adjoint positive
{\it pseudo}-differential operator of first order.

In Sec. 2 we motivate the study of the invariants: 
the {\it relativistic heat 
trace}
\bea
\Theta_r(\beta) &=& \Tr \exp(-\beta\omega)
\eea
and the {\it quantum heat 
traces}
\bea
\Theta_b(\beta,\mu) &=&\Tr \left\{\exp[\beta(\omega-\mu)]-1\right\}^{-1},
\\[5pt]
\Theta_f(\beta,\mu) &=&\Tr \left\{\exp[\beta(\omega-\mu)]+1\right\}^{-1},
\eea
where $\Tr$ denotes the standard $L^2$ trace, $\beta$ is a positive
parameter and $\mu$ is a (generally, non-positive) parameter. 
We also introduce the corresponding zeta functions: 
the {\it relativistic zeta function}
\bea
Z_{r}(s,\mu) &=&
\frac{1}{\Gamma(s)}\int_0^\infty d\beta \; \beta^{s-1}
e^{\beta\mu}\Theta_{r}(\beta)
\label{325bbc}
\eea
and the
{\it quantum zeta
functions} 
\bea
Z_{b,f}(s,\mu) &=&
\frac{1}{\Gamma(s)}\int_0^\infty d\beta \; \beta^{s-1}
\Theta_{b,f}(\beta,\mu).
\eea
We show that these new
invariants can be reduced to some integrals of the well known classical heat
trace
\be
\Theta(t)=\Tr \exp(-tH)
\ee
and compute the asymptotics of these invariants as $\beta\to 0$.

In Sec. 3 we review the standard theory of the heat kernel 
of the Laplace type operator $H$ in a form suitable for our analysis.
To be precise, we consider a closed Riemannian manifold $M$ of dimension $n$,
a vector bundle $\cV$ over $M$ and
an elliptic self-adjoint second-order {\it partial} differential operator
$H$
with a positive definite scalar leading symbol of Laplace type
acting on sections of the bundle $\cV$.
We introduce a function $A_q$ of a complex variable $q$ defined by
\be
A_q=(4\pi)^{n/2}\frac{1}{\Gamma(-q)}\int_0^\infty
dt\;t^{-q-1+n/2}\Theta(t).
\ee
Then we show that for a positive operator $H$ the function $A_q$
is entire and its values
at non-negative integer points $q=k$ are equal to the standard heat trace
coefficients, which are locally computable, while the values
of the function $A_q$ at half-integer points $q=k+1/2$ as well
as the values of its derivative at the positive integer points
are new global invariants that are not locally computable.

In Sec. 4 we compute the relativistic zeta function $Z_r(s,\mu)$
and the asymptotics of the relativistic heat trace $\Theta_r(\beta)$
as $\beta\to 0$.
We obtained
in even dimension $n=2m$,
\bea
\Theta_{r}(\beta)&\sim& 
\sum_{k=0}^\infty \beta^{2k-2m}b^{(1)}_k A_k
+\sum_{k=0}^\infty \beta^{2k+1}b^{(2)}_{k}A_{k+m+1/2},
\eea
and in odd dimension $n=2m+1$,
\bea
\Theta_{r}(\beta)&\sim& 
\sum_{k=0}^\infty\beta^{2k-2m-1}
b^{(3)}_kA_k
+\log\beta\sum_{k=0}^\infty 
\beta^{2k+1}b^{(4)}_{k}A_{k+m+1}
+\sum_{k=0}^\infty \beta^{2k+1}
b^{(5)}_{k}A'_{k+m+1},
\nonumber\\
\eea
and computed all numerical coefficients $b^{(i)}_k$.

In Secs. 5 and 6 we express the quantum heat traces in terms of the
classical one
and compute their asymptotics as $\beta\to 0$. 
For 
$\mu=0$ we obtain an asymptotic expansion as $\beta\to 0$:
in even dimension $n=2m$,
\bea
\Theta_{f}(\beta,0)&\sim& 
\sum_{k=0}^m \beta^{2k-2m}c^{(1)}_k A_k
+\sum_{k=0}^\infty \beta^{2k+1}c^{(2)}_{k}A_{k+m+1/2},
\\
\Theta_{b}(\beta,0) &=&
\sum_{k=0}^m\beta^{2k-2m}c^{(3)}_kA_k
+\sum_{k=-1}^\infty \beta^{2k+1}c^{(4)}_{k}A_{k+m+1/2},
\eea
and in odd dimension $n=2m+1$,
\bea
\Theta_{f}(\beta,0)&\sim& 
\sum_{k=0}^\infty\beta^{2k-2m-1}
c^{(5)}_kA_k
+\log\beta\sum_{k=0}^\infty 
\beta^{2k+1}c^{(6)}_{k}A_{k+m+1}
+\sum_{k=0}^\infty \beta^{2k+1}
c^{(7)}_{k}A'_{k+m+1},
\nonumber\\
\\
\Theta_{b}(\beta,0) &=&
\sum_{k=0}^{m-1} 
\beta^{2k-2m-1}
c^{(8)}_kA_k
+\log\beta\sum_{k= -1}^\infty 
\beta^{2k+1}
c^{(9)}_{k}A_{k+m+1}
+\sum_{k=-1}^\infty 
\beta^{2k+1}c^{(10)}_{k}A'_{k+m+1}.
\nonumber\\
\eea
and computed all numerical coefficients $c^{(i)}_k$.

%=====================================
\section{Quantum Heat Traces}
\setcounter{equation}0

%\subsection{Statistical Physics}

We are going to introduce some new invariants of elliptic operators on manifolds.
We will draw a deep analogy from physics. 
In statistical physics \cite{landau69}
one considers a statistical ensemble of $N$ {\it
identical} particles. The particles can be in states with a discrete set of
energy levels $\{\omega_k\}_{k=1}^\infty$ bounded below, $\omega_k\ge
\omega_1$; we assume that the energy levels form an increasing unbounded
sequence of real numbers. The state with the lowest energy $\omega_1$ is called
the {\it ground (or vacuum) state}. Without loss of generality one can always assume
that the vacuum energy is positive, i.e. $\omega_1>0$.

Let $n_k$ be the average number of particles in the state with the energy
$\omega_k$. If the ensemble is in 
{\it thermodynamical equilibrium} then the number
of particles $n_k$ with the energy $\omega_k$ is determined by just two parameters:
the temperature $T$ and the chemical potential $\mu$. In a particular type of
systems, such as photon gas, when the number of particles is not conserved the
chemical potential is exactly equal to zero, $\mu=0$. In classical physics,
when the particles are {\it distinguishable}, this number is determined by the
{\it Boltzman distribution}
\be
n_k=\exp[-\beta (\omega_k-\mu)],
\ee
where, following standard notation, $\beta=1/T$ is the inverse temperature.

Of course, the total number of particles is
\be
N(\beta,\mu)=\sum_{k=1}^\infty\exp[-\beta (\omega_k-\mu)],
\ee
which is usually taken as the implicit definition of the chemical potential
as a function of the temperature and the number of particles
%$\mu=\mu(\beta,N)$ .
\be
\mu(\beta,N)=-\frac{1}{\beta}\log\left\{\frac{1}{N}
\sum_{k=1}^\infty\exp(-\beta \omega_k)\right\}.
\ee

The energy of a 
non-relativistic classical particle of mass $m$ is described by
the {\it classical Hamiltonian}
\be
\omega_c=\frac{1}{2m}H_0,
\ee
where $H_0$ is an elliptic self-adjoint second-order partial differential
operator with positive definite leading symbol of Laplace type
acting on sections of a vector bundle over a closed
manifold of dimension $n$.
The eigenvalues $\lambda_k$ of the operator $H_0$ play the role of the square of momenta;
they are bounded from below.
Thus, the total number of classical particles is given by 
\be
\Tr \exp[-\beta(\omega_c-\mu)]
=e^{\beta\mu}
\Theta_0\left(\frac{\beta}{2m}\right),
\ee
where
\be
\Theta_0(t)=\Tr\exp\left(-tH_0\right)
\label{214zz}
\ee
is the
well-known {\it classical heat trace}.

The energy of a relativistic classical particle is described by an elliptic self-adjoint
first-order pseudo-differential operator
(that we call the {\it relativistic Hamiltonian})
\be
\omega=H^{1/2},
\label{29xx}
\ee
where
\be
H=H_0+m^2.
\ee
In order for this to make sense we will always assume that the mass parameter is
large enough so that the operator $H$ is positive.
Of course, particles with low momenta and large mass, when $m^2>>\lambda_k$,
are non-relativistic with the energy
 \footnote{Strictly
speaking, this Hamiltonian only describes the energy levels of the particles
with low momenta when the eigenvalues of the operator $H_0$ are smaller than
$m^2$.}
\be
m+\omega_c=m+\frac{1}{2m}H_0.
\ee
Therefore,  the total number of relativistic classical particles is given by
\be
\Tr \exp[-\beta(\omega-\mu)]
=e^{\beta\mu}
\Theta_r\left(\beta\right),
\ee
where
\be
\Theta_r(t)=\Tr\exp\left(-t\omega\right).
\label{214zzbb}
\ee
is the
trace of an elliptic pseudo-differential positive operator of first order
(that we call {\it relativistic heat trace}).

Contrary to the classical case, quantum particles are {\it indistinguishable}.
As a matter of fact, there are two kinds of quantum particles, {\it bosons} and
{\it fermions}. The number of bosons with energy $\omega_k$ is given by the
{\it Bose-Einstein distribution}
\be
n_{b,k}=\frac{1}{\exp[\beta (\omega_k-\mu)]-1}.
\ee
Notice also that this distribution function is well defined for all energies
only if $\mu<\omega_1$ since $n_{b,1}$ diverges as $\mu\to \omega_1$.
%\be
%n_b(E_0)\sim \frac{1}{\beta(E_0-\mu)}.
%\ee
Obviously, particles with large energies, more precisely, when
$\exp[\beta(\omega_k-\mu)]>>1$, are still distributed according to the
classical Boltzman distribution.

The total number of particles is now
\be
N_b(\beta,\mu) = \sum_{k=1}^\infty \frac{1}{\exp[\beta (\omega_k-\mu)]-1}.
\label{24xx}
\ee
This again implicitly defines the chemical potential $\mu=\mu(\beta,N)$ as a
function of the temperature and the number of particles. Of course, if we
{\it fix the number of particles} $N$ and vary the temperature $\beta$ then the
chemical potential is only a function of the temperature, $\mu=\mu(\beta)$. One
can show \cite{landau69} that in this case there must exist a critical
temperature $\beta_c$ such that at large temperature, $\beta<\beta_c$, the
number of particles in the vacuum, $n_{b,1}(\beta)$, is negligible. On another
hand, at low temperatures, $\beta>\beta_c$, the chemical potential $\mu(\beta)$
approaches $\omega_1$ with an increasing number of particles in the ground
state. In the limit of zero temperature, $\beta\to \infty$, all particles will
be in the ground state. The critical temperature $\beta_c$ is implicitly
defined by the equation
\be
N_b = \sum_{k=2}^\infty \frac{1}{\exp[\beta_c (\omega_k-\omega_1)]-1},
\ee
where the summation goes over all states except the ground state. This leads to
the so-called {\it Bose-Einstein condensation} of quantum particles in the
ground state with energy $\omega_1$.

The fermions obey the {\it Pauli exclusion principle}, which simply states that no two fermions
can be in the same quantum state. This leads to a different energy distribution, so called
{\it Fermi-Dirac distribution}
\be
n_{f,k} = \frac{1}{\exp[\beta (\omega_k-\mu)]+1}.
\ee
Obviously, the Fermi-Dirac distribution also becomes the Boltzman distribution
for large energy particles. 
Contrary to the Bose-Einstein distribution the
Fermi-Dirac distribution is defined for any chemical potential. Notice that the
number of fermions $n_{f,k}$ in any state is always less than $1$. It is a
monotonically decreasing function of energy. Moreover, at zero temperature, as
$\beta\to \infty$, the distribution function approaches the step function
\be
n_{f,k}(\beta,\mu)\sim 
\theta(\mu-\omega_k) = 
\left\{
\begin{array}{ll}
1, & \mbox{ if } \omega_k<\mu,
\\[10pt]
\displaystyle\frac{1}{2}, & \mbox{ if } \omega_k=\mu,
\\[10pt]
0, & \mbox{ if } \omega_k>\mu.
\end{array}
\right.
\ee
The total number of particles is now
\be
N_f(\beta,\mu)=\sum_{k=0}^\infty \frac{1}{\exp[\beta (\omega_k-\mu))]+1}.
\ee
Therefore, at zero temperature the total number of fermions  is simply equal to the
number ${\cal N}(\mu)$ of quantum states with energy less than $\mu$, that is, as $\beta\to \infty$,
\be
N_f(\beta,\mu)\sim {\cal N}(\mu).
\ee

%=====================================
%\subsection{Quantum Heat Traces}
%\setcounter{equation}0

Thus, for quantum particles we are led to define two types of new
invariants, that we call {\it quantum heat traces}, the bosonic and the
fermionic one by
\bea
\Theta_b(\beta,\mu) &=&\Tr \left\{\exp[\beta(\omega-\mu)]-1\right\}^{-1},
\label{219vv}
\\[10pt]
\Theta_f(\beta,\mu) &=&\Tr \left\{\exp[\beta(\omega-\mu)]+1\right\}^{-1},
\label{220vv}
\eea
where $\omega$ is the relativistic Hamiltonian (\ref{29xx}).

Here the parameter $\mu$ is essential and cannot be simply factored out.
In this paper we will restrict mostly to the case of
positive operator $\omega$, that is, $\omega_1>0$.
We will assume that in the bosonic case the chemical potential
$\mu$ is non-positive,
$\mu\le 0$ and in the fermionic case it can
take any real values. 
For the fermionic case with $\mu>0$ we will decompose the operator as follows
\be
\Theta_f(\beta,\mu)=\sum_{\omega_k\le\mu}
\frac{1}{\exp[\beta(\omega_k-\mu)]+1}
+\Tr (I-P_\mu)\left\{\exp[\beta(\omega-\mu)]+1\right\}^{-1},
\ee
where $P_\mu$ is the projection operator on the bottom of the spectrum
with eigenvalues less or equal to $\mu$.
Note that
\be
\Tr (I-P_\mu)\exp(-tH)=
\Theta(t)-\sum_{\omega_k\le\mu}\exp(-t\omega_k^2).
\ee

Note that at zero temperature, as $\beta\to \infty$, the fermionic heat trace
becomes simply the {\it spectral counting function} (up to exponetially small
terms)
\be
\Theta_f(\beta,\mu)\sim {\cal N}(\mu),
\ee
where ${\cal N}(\mu)$ is the number of eigenvalues of the operator $\omega$
less or equal than $\mu$ 
(when $\mu$ coincides with an eigenvalue there is
a correction term $1/2$). 
Therefore, it can be considered as the regularized version of
the spectral counting function.

It is easy to see that when $\exp[\beta(\omega_1-\mu)]>>1$, that is, for
particles with large energies, or, more precisely,
as $\mu\to-\infty$, both quantum heat traces are determined by the
relativistic heat trace
\be
\Theta_b(\beta,\mu) \sim 
\Theta_f(\beta,\mu) \sim e^{\beta\mu}\Theta_r(\beta).
\ee

One can go even further and define,
via the Mellin transform, the corresponding 
{\it relativistic zeta function}
\bea
Z_{r}(s,\mu) &=&
\frac{1}{\Gamma(s)}\int_0^\infty d\beta \; \beta^{s-1}
e^{\beta\mu}\Theta_{r}(\beta)
\label{325bbb}
\eea
and the
{\it quantum zeta
functions} 
\bea
Z_{b,f}(s,\mu) &=&
\frac{1}{\Gamma(s)}\int_0^\infty d\beta \; \beta^{s-1}
\Theta_{b,f}(\beta,\mu).
\eea
It is not difficult to show that for non-positive $\mu\le 0$
\bea
Z_{r}(s,\mu) &=&\Tr (\omega-\mu)^{-s}
\eea
and 
\bea
Z_{b}(s,\mu) &=&
\zeta(s)Z_r(s,\mu)\,,
\label{328bbb}
\\
Z_{f}(s,\mu) &=&
(1-2^{1-s})\zeta(s)Z_r(s,\mu)
\,.
\label{329bbb}
\eea
%In the limit of large mass we get the standard spectral zeta function
%\be
%\Tr(\omega-\mu)^{-s}=(2m)^{-s}\Tr\left[H+2m(m-\mu)\right]^{-s}.
%\ee

The analyticity properties of these zeta functions depend on the asymptotics of
the quantum heat traces as $\beta\to 0$ and $\beta\to \infty$, which, in turn,
depend crucially on the chemical potential $\mu$. 
It is not difficult to see that for non-positive chemical potential
these zeta functions are analytic for 
sufficiently large real part of $s$.
More precisely, as shown below, as $\beta\to 0$ the quantum heat traces behave as
$\beta^{-n}$, with $n=\dim M$, 
and, therefore, the quantum zeta functions are analytic for
$\Real s>n$.
Then by using the inverse Mellin
transform we can express the quantum heat traces in terms of the zeta functions
\bea
\Theta_{b,f}(\beta,\mu)=\frac{1}{2\pi i}\int_{c-i\infty}^{c+i\infty} ds\;
\beta^{-s}\Gamma(s)Z_{b,f}(s,\mu)\,,
\label{330bbb}
\eea
where $c>n$. 

%=====================================
\section{Heat Kernel of Laplace Type Operator}
\setcounter{equation}0

We are going to employ a very useful representation of the classical heat
kernel developed in \cite{avramidi91} (see also
\cite{avramidi00,avramidi10,avramidi98,avramidi99,avramidi15}).

%%%%%%%%%%%%%%%%%%%%%%%%%%%%%%%%%%%%%%%%%%%%%%%%%%

%=====================================================
%\subsection{Laplace Type Operators}

Let $(M,g)$ be a smooth compact Riemannian manifold of dimension $n$ without
boundary, equipped with a positive definite Riemannian metric $g$.
%We assume that it is complete simply connected orientable and spin.
We denote the local coordinates on $M$ by $x^\mu$, with Greek indices running
over $1,\dots, n$. The Riemannian volume element is defined as usual by
$d\vol=dx\,g^{1/2}\,,$ where $g=\det g_{\mu\nu}\,.$ Let ${\cal V}$ be a vector
bundle over the manifold $M$, $\nabla$ be a connection and ${\cal A}_\mu$ be
the connection one-form on the bundle $\mathcal{V}$. We assume that the vector
bundle ${\cal V}$ is equipped with a Hermitian metric. This naturally
identifies the dual vector bundle ${\cal V}^*$ with ${\cal V}$. We assume that
the connection $\nabla$ is compatible with the Hermitian metric on the vector
bundle ${\cal V}$. The connection is given its unique natural extension to
bundles in the tensor algebra over ${\cal V}$ and ${\cal V}^*$, and, using the
Levi-Civita connection of the metric $g$, to all bundles in the tensor algebra
over ${\cal V},\,{\cal V}^*,\,TM$ and $T^*M$; the resulting connection will
usually be denoted just by $\nabla$.

We denote by $C^\infty({\cal V})$ the space of smooth sections of the bundle
${\cal V}$. The fiber inner product $\left<\;,\;\right>$ on the bundle ${\cal
V}$ and the fiber trace, $\tr$, defines a natural $L^2$ inner product $(\;,\;)$
and the $L^2$-trace, $\Tr$, using the invariant Riemannian measure on the
manifold $M$. The completion of $C^\infty({\cal V})$ in this norm defines the
Hilbert space $L^2({\cal V})$ of square integrable sections.

Let $\nabla^*$ be the formal adjoint to $\nabla$ defined using the Riemannian
metric and the Hermitian structure on ${\cal V}$ and let $Q$ be a smooth
Hermitian section of the endomorphism bundle $\End({\cal V})$. A Laplace type
operator $H: C^\infty(V)\to C^\infty(V)$ is a partial differential operator of
the form
\be 
H=\nabla^*\nabla+Q=-\Delta+Q\,,
\label{1ms}
\ee
where $\Delta$ is the Laplacian which in local coordinates takes the form
\be
\Delta=g^{\mu\nu}\nabla_\mu\nabla_\nu=
g^{-1/2}(\partial_\mu+{\cal A}_\mu)g^{1/2}g^{\mu\nu}
(\partial_\nu+{\cal A}_\nu).
\ee
Thus, a Laplace type operator is constructed from the following three pieces of
geometric data: i) a Riemannian metric $g$ on $M$, ii) a connection $1$-form
${\cal A}$ on the vector bundle ${\cal V}$ and iii) an endomorphism $Q$ of the
vector bundle ${\cal V}$. It is not difficult to show that every elliptic
second-order partial differential operator with a positive definite scalar
leading symbol is of Laplace type and can be put in this form by choosing the
appropriate metric, connection and the endomorphism $Q$. It is easy to show
that the operator $H$ is an elliptic essentially self-adjoint partial
differential operator. It is well known \cite{gilkey95} that the spectrum of
the operator $H$ has the following properties: i) the spectrum is discrete,
real and bounded from below, ii) the eigenspaces are finite-dimensional, and
iii) the eigenvectors are smooth sections of the vector bundle ${\cal V}$ that
form a complete orthonormal basis in $L^2({\cal V})$. We denote the eigenvalues
and the eigenvectors of the operator $H$ by $\{\lambda_k,\varphi_k\}_{k\in
\ZZ_+}$.

%===============================================
%\subsection{Spectral Functions}
%=============================================
%\subsection{Heat Kernel}

The heat semi-group is the family of bounded operators on $L^2({\cal V})$ for
$t>0$
\be
U(t)=\exp(-tH)\,.
\ee
The kernel of this operator (called the {\it heat kernel}) is defined by
\be 
U(t;x,x')
=\sum\limits_{k=1}^\infty
e^{-t\lambda_k}\varphi_k(x)\otimes \varphi^*_k(x'),
\ee
where each eigenvalue is counted with multiplicity;
it satisfies the heat equation 
\be
\left(\partial_t+H\right)U(t)=0
\label{he-5/01}
\ee
with the initial condition
\be
U(0^+;x,x')=\delta(x,x')\,.
\label{init-5/01}
\ee
The heat trace 
can be expressed as the integral of the heat kernel diagonal
\be
\Theta(t)=
\Tr\exp(-tH)=\int\limits_M d\vol\; \tr U(t;x,x)\,.
\ee
%hereafter $\tr$ denotes the fiber trace.

That is why we are going to study the heat kernel $U(t;x,x')$ in the
neighborhood of the diagonal of $M\times M$, when the points $x$ and $x'$ are
close to each other. We will keep a point $x'$ of the manifold fixed and
consider a small geodesic ball, i.e. a small neighborhood of the point $x'$:
$B_\varepsilon(x') =\{x\in M| r(x,x')<\varepsilon\}$, $r(x,x')$ being the
geodesic distance between the points $x$ and $x'$. We will take the radius of
the ball sufficiently small, so that each point $x$ of the ball of this
neighbourhood can be connected by a unique geodesic with the point $x'$. This
can be always done if the size of the ball is smaller than the injectivity
radius of the manifold, $\varepsilon<r_{\rm inj}$. Let $\sigma(x,x')$ be the
geodetic interval, also called world function, defined as one half the square
of the length of the geodesic connecting the points $x$ and $x'$
\be
\sigma(x,x')=\frac{1}{2}r^2(x,x').
\ee
The first derivatives of this function with respect to $x$ and $x'$,
$\sigma_\mu=\nabla_\mu \sigma$ and $\sigma_{\mu'}=\nabla'_{\mu'}\sigma$, define
tangent vector fields to the geodesic at the points $x$ and $x'$ respectively
pointing in opposite directions and the determinant of the mixed second
derivatives defines the so called Van Vleck-Morette determinant
\be
\Delta(x,x')=g^{-1/2}(x)
\det\left[-\nabla_\mu\nabla'_{\nu'}\sigma(x,x')\right]g^{-1/2}(x').
\ee
This object should not be confused with the Laplacian. Let, finally, ${\cal
P}(x,x')$ denote the parallel transport operator of sections of the vector
bundle ${\cal V}$ along the geodesic from the point $x'$ to the point $x$. Here
and everywhere below the coordinate indices of the tangent space at the point
$x'$ are denoted by primed Greek letters. They are raised and lowered by the
metric tensor $g_{\mu'\nu'}(x')$ at the point $x'$. The derivatives with
respect to $x'$ will be denoted by primed Greek indices as well.

%==========================================
%\subsubsection{Normal Coordinates}
%======================
%===========================================
%\subsubsection{Transport Function}

To study the local behavior of the heat kernel we use the following
ansatz (motivated by the heat kernel in the Euclidean space)
\be
U(t;x,x')=(4\pi t)^{-n/2}\Delta^{1/2}(x,x')
\exp\left(-\frac{1}{2t}\sigma(x,x')\right){\cal P}(x,x')
\Omega(t;x,x').
\label{150}
\ee
The function $\Omega(t;x,x')$, called the {\it transport function}, is a
section of the endomorphism vector bundle $\End(V)$ over the point $x'$. For
the sake of simplicity we will omit the space variables $x$ and $x'$ when it
will not cause any confusion. Using the definition of the functions
$\sigma(x,x')$, $\Delta(x,x')$ and ${\cal P}(x,x')$ it is not difficult to find
that the transport function satisfies a transport equation
\be
\left(\partial_t+\frac{1}{t}D+\tilde H\right)\Omega(t)=0,
\ee
where $D$ is the radial vector field, i.e. operator of differentiation along
the geodesic,
defined by
\be
D=\sigma^\mu\nabla_\mu,
\ee
and $\tilde H$ is a second-order differential operator defined by
(here, again, $\Delta$ is not the Laplacian bu the Van-Vleck-Morette
determinant)
\be
\tilde H={\cal P}^{-1}\Delta^{-1/2}H\Delta^{1/2}{\cal P}.
\label{160}
\ee
The initial condition for the transport function is obviously
\be
\Omega(0;x,x')=I,
\ee
where $I$ is the identity endomorphism of the vector bundle ${\cal V}$ over
$x'$. One can show that if the operator $H$ is positive then the transport
function $\Omega(t)$ satisfies the following asymptotic conditions
\be
\lim_{t\to\infty,0}t^\alpha\partial_t^N\Omega(t)=0
\qquad
\mbox{for any}\ \alpha\in \RR_+, \ N\in \ZZ_+.
\label{200}
\ee

Now, we consider a slightly modified version of the Mellin transform of the
function $\Omega(t)$ introduced in \cite{avramidi91}
\be
a_q=\frac{1}{\Gamma(-q)}
\int\limits_0^\infty dt\;
t^{-q-1}\Omega(t).
\label{41-7/98}
\ee
The integral (\ref{41-7/98}) converges for ${\rm Re}\, q<0$. By integrating by
parts $N$ times and using the asymptotic conditions (\ref{200}) we also get
\be
a_q
=\frac{1}{\Gamma(-q+N)}\int\limits_0^\infty dt\;
t^{-q-1+N}(-\partial_t)^N
\Omega(t).
\ee
This integral converges for ${\rm Re}\,q<N-1$. Using this representation one
can prove that \cite{avramidi91} the function $a_q$ is an entire function of
$q$ satisfying the asymptotic condition
\be
\lim_{|q|\to\infty,\ {\rm Re}\,q<N}\Gamma(-q+N)a_q=0,\qquad
{\rm for\ any}\ N>0.
\label{350}
\ee
Moreover, 
the values of the function $a_q$ at the integer positive points
$q=k$ are given by the Taylor coefficients
\be
a_k=(-\partial_t)^k\Omega(t)\Big|_{t=0}.
\label{43-7/98}
\ee

By inverting the Mellin transform we obtain for the transport
function
\be
\Omega(t)=\frac{1}{2\pi i}\int\limits_{c-i\infty}^{c+i\infty}dq\;
t^q\,\Gamma(-q)a_q
\label{46-7/98}
\ee
where $c<0$, which, by using our ansatz (\ref{150}),
immediately gives also the heat trace
\be
\Theta(t)=(4\pi)^{-n/2}
\frac{1}{2\pi i}
\int\limits_{c-i\infty}^{c+i\infty}dq\,t^{q-n/2}\,\Gamma(-q)A_q,
\label{421bb}
\ee
where
\be
A_q=\int\limits_M d\vol\;\tr\,a_q(x,x).
\ee
Note that the global function $A_q$ is determined by the 
Mellin transform of the heat trace
\be
A_q=(4\pi)^{n/2}\frac{1}{\Gamma(-q)}
\int\limits_0^\infty dt\;
t^{-q-1+n/2}\Theta(t).
\label{223bb}
\ee

Substituting this ansatz into the transport equation we get a functional
equation for the function $a_q$
\be
\left(1+\frac{}{q}D\right)a_q
=\tilde H\,a_{q-1}.
\label{400-7/98}
\ee
The initial condition for the transport function is translated into
\be
a_0=I.
\label{500-7/98}
\ee
{}For integer $q=k=1,2,\dots$ the functional equation (\ref{400-7/98}) becomes
a recursion system that, together with the initial condition (\ref{500-7/98}),
determines all coefficients $a_k$.

Thus, we have reduced the problem of solving the heat equation to the following
problem: one has to find an entire function of $q$, $a_q(x,x')$, that satisfies
the functional equation (\ref{400-7/98}) with the initial condition
(\ref{500-7/98}) and the asymptotic condition (\ref{350}). The function $a_q$
turns out to be extremely useful in computing the heat kernel, the resolvent
kernel, the zeta-function and the determinant of the operator $H$. It contains
the same information about the operator $H$ as the heat kernel. In some cases
the function $a_q$ can be constructed just by analytical continuation from the
integer positive values $a_k$ \cite{avramidi91}. Now we are going to do the
usual trick, namely, to move the contour of integration over $q$ in
(\ref{46-7/98}) to the right. Due to the presence of the gamma function
$\Gamma(-q)$ the integrand has simple poles at the non-negative integer points
$q=0,1,2\dots$, which contribute to the integral while moving the contour. This
gives the asymptotic expansion as $t\to 0$
\be
\Omega(t)\sim 
\sum\limits_{k=0}^\infty\frac{(-t)^k}{k!}a_k.
\label{600}
\ee
The heat trace has an analogous asymptotic expansion as
$t\to 0$
\be
\Theta(t)\sim (4\pi t)^{-n/2}
\sum\limits_{k=0}^\infty \frac{(-t)^k}{k!}A_k
\,,
\ee
where
\be
A_k=\int\limits_M d\vol\;\tr\,a_k(x,x).
\ee

%=======================================================
%\subsection{Zeta Function and Determinant}

We can apply our ansatz for computation of the complex power of the
operator $H$ defined
by
\be
G_s(x,x')=\frac{1}{\Gamma(s)}
\int\limits_0^\infty dt\, t^{s-1}\,\,U(t;x,x').
\label{428bb}
\ee
Outside the diagonal, i.e. for $x\ne x'$ or $\sigma\ne 0$, this integral
converges for any $s$ and defines an entire function of $s$. Using our ansatz
for the heat kernel one can obtain the so-called {\it Mellin-Barnes representation} of the
function $G_s$ \cite{avramidi91}
\be
G_s(x,x')
=(4\pi)^{-n/2}\Delta^{1/2}{\cal P}
\frac{1}{2\pi i}\int\limits_{c-i\infty}^{c+i\infty}
dq\,\frac{\Gamma(-q)\Gamma(-q-s+n/2)}{\Gamma(s)}
\left(\frac{\sigma}{2}\right)^{q+s-n/2}a_q(x,x')
\label{429bb}
\ee
where $c<-{\rm Re}\,s+n/2$. The integrand in this formula is a meromorphic
function of $q$ with some simple and maybe some double poles. If we move the
contour of integration to the right, we get contributions from the simple poles
in form of powers of $\sigma$ and a logarithmic part due to the double poles
(if any). This gives the complete structure of diagonal singularities of
$G_s(x,x')$ (see \cite{avramidi98}).

For sufficiently large $\Real s$, more precisely, $\Real s>n/2$ the integral
(\ref{428bb}) converges even on the diagonal, that is, for $x=x'$. In this case
there are no singularities at all and there is a well defined diagonal limit as
$x\to x'$
\be
G_s(x,x)=\sum_{k=1}^\infty \frac{1}{\lambda_k^s}\varphi_k(x)\varphi^*(x).
\ee
On another hand, by using (\ref{41-7/98}) on the diagonal we obtain
\be
a_q(x,x)=(4\pi )^{n/2}\frac{\Gamma(-q+n/2)}{\Gamma(-q)}
\sum_{k=1}^\infty \frac{1}{\lambda_k^{-q+n/2}}\varphi_k(x)\varphi^*(x);
\ee
therefore, we obtain a very simple formula
\be
G_s(x,x)
=(4\pi)^{-n/2}\frac{\Gamma(s-n/2)}{\Gamma(s)}a_{n/2-s}(x,x).
\ee

This gives automatically the zeta-function 
\be
\zeta_H(s)=\Tr H^{-s}=(4\pi)^{-n/2}
\frac{\Gamma(s-n/2)}{\Gamma(s)}A_{n/2-s}.
\label{234bbb}
\ee
Since $A_q$ is an entire function this shows that the zeta function is a
meromorphic function and gives the complete structure of its singularities
(simple poles): an infinite sequence at the half-integer points
$s=[n/2]+1/2-k$, $(k=0,1,2,\dots)$ for odd dimension and a finite number 
of poles at
integer points $s=1,2,\dots, n/2$ for even dimension. In particular, the
zeta-function is analytic at the origin; its value at the origin is given by
\be
\zeta_H(0)=\left\{
\begin{array}{ll}
0 & \mbox{ for odd } n=2m+1\,,\\
(4\pi)^{-m}{\displaystyle \frac{(-1)^{m}}{m!}}
A_{m} & \mbox{ for even } n=2m\,,
\end{array}
\right.
\ee
and its derivative at the origin (which determines the determinant
of the operator $H$) is
\be
\zeta'_H(0)=\left\{
\begin{array}{ll}
\displaystyle
(-1)^{m+1}\pi^{-m}\frac{m!}{(2m+1)!}A_{m+1/2}
& \mbox{ for odd } n=2m+1\,,
\\[14pt]
\displaystyle
(4\pi)^{-m}\frac{(-1)^{m}}{m!}
\left\{-A'_{m}+[\psi(m+1)+{\bf C}]A_{m}
\right\}
& \mbox{ for even } n=2m\,.
\end{array}
\right.
\ee
Here $\psi(z)=(d/dz)\log\Gamma(z)$ is the psi-function,
${\bf C}=-\psi(1)=0.577\dots$ is
the Euler constant, and
\be
A'_{m}=\frac{\partial}{\partial q}
A_q\Big|_{q=m}.
\ee
Note that the value $\zeta_H(0)$ is determined by the locally computable 
heat kernel coefficient $A_k$ whereas the derivative of the zeta function 
$\zeta'_H(0)$ is determined by the global invariants $A_{m+1/2}$
and $A'_{m}$ which are not
locally computable.

%%%%%%%%%%%%%%%%%%%%%%%%%%%%%%%%%%%%%%%%%%%%%%%%%%
%=====================================
\section{Relativistic Heat Trace and Zeta Function}
\setcounter{equation}0

First of all, we compute the relativistic zeta function. 
We use the integral
\cite{prudnikov83}
\be
\exp(-x)=
(4\pi)^{-1/2}
\int_0^\infty
dt\;t^{-3/2}\exp\left(-\frac{1}{4t}-tx^2\right),
\label{31bb}
\ee
which is valid for any $x\ge 0$ and eq. (\ref{325bbb})
to obtain
\be
Z_r(s,\mu)=\int_0^\infty dt\; g(t,s,\mu)\Theta(t),
\ee
where
\be
g(t,s,\mu)=(4\pi)^{-1/2}\frac{1}{\Gamma(s)}t^{-3/2}
\int_0^\infty d\tau\; \tau^{s}\exp\left(-\frac{\tau^2}{4t}+\mu\tau\right).
\ee
Now, by expanding this in powers of $\mu$ we get
\be
g(t,s,\mu)=(4\pi)^{-1/2}\frac{1}{\Gamma(s)}
\sum_{k=0}^\infty \frac{\mu^k}{k!}t^{(s+k)/2-1}
2^{s+k}\Gamma[(s+k+1)/2].
\ee
Further, by using the obvious relation (\ref{223bb})
we obtain
\be
Z_r(s,\mu)=(4\pi)^{-(n+1)/2}\frac{1}{\Gamma(s)}
\sum_{k=0}^\infty \frac{\mu^k}{k!}
2^{s+k}\Gamma[(s+k+1)/2]\Gamma[(s+k-n)/2]
A_{(n-s-k)/2}.
\ee
This enables one to extend the zeta function $Z_r(s,\mu)$ to a meromorphic function of $s$,
in particular, it is analytic at $s=0$.
Note that the same result can be obtained by using the expansion
\be
Z_r(s,\mu)=\sum_{k=0}^\infty \frac{\mu^k}{k!}\frac{\Gamma(s+k)}{\Gamma(s)}
\zeta_H[(s+k)/2]
\ee
and the eq. (\ref{234bbb}).
As a result, for $\mu=0$ we obtain a very simple formula
\be
Z_r(s,0)=(4\pi)^{-(n+1)/2}2^s\frac{\Gamma[(s+1)/2]\Gamma[(s-n)/2]}{\Gamma(s)}
A_{(n-s)/2}.
\label{47bbb}
\ee

Next, we use the integral (\ref{31bb})
to reduce the relativistic heat trace to the 
classical heat trace (\ref{214zz}), that is,
\be
\Theta_r(\beta)=
(4\pi)^{-1/2}
\int_0^\infty
dt\;t^{-3/2}\exp\left(-\frac{1}{4t}\right)\Theta(t\beta^2).
\label{32bb}
\ee
Now, by using eqs. (\ref{32bb}) and (\ref{421bb})  we can express
the relativistic heat trace in terms of the function $A_q$
via a Mellin-Barnes integral
\be
\Theta_r(\beta)=
2(4\pi)^{-(n+1)/2}
\frac{1}{2\pi i}
\int\limits_{c-i\infty}^{c+i\infty}dq\,\Gamma(-q)\Gamma\left[-q+(n+1)/2\right]
\left(\frac{\beta}{2}\right)^{2q-n}A_q,
\label{597a}
\ee
It is worth nothing the striking resemblance of this equation to the kernel of
the function $G_s$ given by eq. (\ref{429bb}) by replacing $\sigma$ by
$\beta^2$. Therefore, the asymptotics of the relativistic heat kernel
$\Theta_r(\beta)$ as $\beta \to 0$ can be computed exactly in the same way as
the diagonal singularities of $G_s(x,x')$ as $\sigma\to 0$. Namely, we move the
contour of integration to the right to get contributions from the simple poles
in form of powers of $\beta$ and a logarithmic part, $\log\beta$, due to the
double poles (if any, depending on the dimension).

Integrals of this type are a particular case of the so-called Mellin-Barnes
integrals. They are a very powerful tool in computing the heat trace asymptotics.
Since we will use them quite often we prove the following lemma 

\begin{lemma}
Let $f(q)$ be a function of a complex variable $q$ that is analytic in the right
half-plane and decreases sufficiently fast at infinity in the right half-plane.
Let $c<0$ and $m$ be a positive integer and $I(t)$ and $J(t)$ be two functions
of a positive real variable $t$ defined by
\bea
I(t)&=& \frac{1}{2\pi i}
\int\limits_{c-i\infty}^{c+i\infty}dq\,\Gamma(-q)\Gamma\left(-q+m+1/2\right)
t^q f(q),
\label{52bb}
\\
J(t)&=&
\frac{1}{2\pi i}
\int\limits_{c-i\infty}^{c+i\infty}dq\,\Gamma(-q)\Gamma\left(-q+m+1\right)
t^q f(q).
\label{53bb}
\eea
Then there are asymptotic expansion as $t\to 0$ 
\bea
I(t)&=&I_1(t)+I_2(t)+I_3(t),
\\
J(t)&=&J_1(t)+J_2(t)+J_3(t)+J_4(t),
\eea
where
\bea
I_1(t)&\sim& 
\sqrt{\pi}
\sum_{k=0}^m (-1)^k\frac{(2m-2k)!}{k!(m-k)!}2^{2k-2m}t^k f(k)
%\nonumber
\\
I_2(t)&\sim&(-1)^m\frac{1}{2}\sqrt{\pi}
\sum_{k=0}^\infty \frac{k!}{(k+m+1)!(2k+1)!}2^{2k+2}t^{k+m+1}f(k+m+1)
%\nonumber
\\
I_3(t)&\sim&
-(-1)^m\frac{1}{2}\sqrt{\pi}
\sum_{k=0}^\infty \frac{(k+m)!}{k!(2k+2m+1)!}2^{2k+2m+2}t^{k+m+1/2}f(k+m+1/2).
\nonumber\\
\eea
\bea
J_1(t)&\sim& 
\sum_{k=0}^m (-1)^k\frac{(m-k)!}{k!}t^k f(k)
%\nonumber
\\
J_2(t)&\sim&
(-1)^m
\sum_{k=0}^\infty \frac{1}{k!(k+m+1)!}t^{k+m+1}
\left[\psi(k+m+2)+\psi(k+1)\right]f(k+m+1)
\nonumber
\\
\\
J_3(t)&\sim& -(-1)^m\log t
\sum_{k=0}^\infty \frac{1}{k!(k+m+1)!}t^{k+m+1}
f(k+m+1)
\\
J_4(t)&\sim& 
-(-1)^m
\sum_{k=0}^\infty \frac{1}{k!(k+m+1)!}t^{k+m+1}
f'(k+m+1).
\eea
\end{lemma}

We would like to stress the difference between different parts of these functions.
The functions $I_1$ and $J_1$ are just polynomials. The functions
$I_2$ and $J_2$ contain only integer powers of $t$ with the coefficients
determined by the values of the function $f$ at integer points.
The function $I_3$ contains {\it  half-integer powers} of $t$  with the coefficients
determined by the values of the function $f$ at {\it half-integer points}
and the function $J_3$ is proportional to the {\it logarithm} $\log t$.
Finally, the coeficients of the function $J_4$ are determined by the {\it derivatives}
of the function $f$ at the integer points.

{\it Proof.}
This lemma can
be proved by the residue calculus.
The integrand for the function $I$ has simple poles at the points the integer points
$q=k$ and the half-integer points $q=k+m+1/2$, with
$(k=0,1,2,\dots)$. The integrand for the function $J$ has simple poles at
$q=0,1,2,\dots, m$ and {\it double} poles at the points $q=k+m+1$.
Now, the lemma is proved by using the formula (for an integer
$k\ge 0$ and $z\to 0$) \cite{erdelyi53}
\be
\Gamma(-k+z)=\frac{(-1)^k}{k!}\left\{
\frac{1}{z}+\psi(k+1)+O(z)\right\}
\ee
and using the properties of the gamma-function.

We apply this lemma to the Mellin-Barnes representation of the relativistic heat trace
(\ref{597a}). We decompose the trace according to
\be
\Theta_r(\beta)=\Theta^{\rm sing}_r(\beta)
+\Theta^{\rm loc}_r(\beta)+\Theta^{\rm non-loc}_r(\beta).
\ee
Now, we need to distinguish the cases of
odd and even dimensions. We obtain:
in even dimension $n=2m$,
\bea
\Theta^{\rm sing}_r(\beta)&\sim& 
\frac{1}{2}
(4\pi)^{-m+1/2}
\sum_{k=0}^m (-1)^k\frac{(2m-2k)!}{k!(m-k)!}\frac{1}{\beta^{2m-2k}}A_k,
%\nonumber
\\
\Theta^{\rm loc}_r(\beta)&\sim&
(-1)^m\frac{1}{2}(4\pi)^{-m}
\sum_{k=0}^\infty \frac{k!}{(k+m+1)!(2k+1)!}\beta^{2k+2}A_{k+m+1},
\nonumber
\\
\\
\Theta^{\rm non-loc}_r(\beta)
&\sim&
-(-1)^m\pi^{-m}
\sum_{k=0}^\infty \frac{(k+m)!}{k!(2k+2m+1)!}\beta^{2k+1}A_{k+m+1/2},
%\nonumber
%\\
\eea
and in odd dimension $n=2m+1$,
\bea
\Theta^{\rm sing}_r(\beta)&\sim& 
2(4\pi)^{-m-1}
\sum_{k=0}^m (-1)^k\frac{(m-k)!}{k!}\left(\frac{2}{\beta}\right)^{2m-2k+1} A_k
%\nonumber
\\
\Theta^{\rm loc}_r(\beta)&\sim&
(-1)^m2(4\pi)^{-m-1}
\sum_{k=0}^\infty \frac{1}{k!(k+m+1)!}
\left(\frac{\beta}{2}\right)^{2k+1}
\nonumber\\
&&\times
\Biggl\{
\psi(k+m+2)+\psi(k+1)
 -2\log(\beta/2)
\Biggr\}A_{k+m+1}
\nonumber\\
\\
\Theta^{\rm non-loc}_r(\beta)&\sim& 
-(-1)^m2(4\pi)^{-m-1}
\sum_{k=0}^\infty \frac{1}{k!(k+m+1)!}\left(\frac{\beta}{2}\right)^{2k+1}
A'_{k+m+1}.
\eea

Note that the coefficients of the parts $\Theta^{\rm sing}_r(\beta)$ and  
$\Theta^{\rm loc}_r(\beta)$ are determined by the heat kernel coefficients $A_k$
(which are locally computable) whereas the part 
$\Theta^{\rm non-loc}_r(\beta)$ is determined by the values of the function
$A_q$ at the half-integer points $A_{k+m+1/2}$ and the derivatives at the integer
points $A'_{k+m+1}$, which are non-locally computable; they are rather related to
the values of the zeta function and its derivatives at negative half-integer points
$\zeta_H(-k-1/2)$ and $\zeta'_H(-k-1/2)$, $k=0,1,2\dots$.

%%%%%%%%%%%%%%%%%%%%%%%%%%%%%%%%%%%%%%%%%%%%%%%%%
%===============================================================

\section{Reduction of Quantum Heat Traces}
\setcounter{equation}0

Next, we will reduce the quantum heat traces to the classical heat trace as well.
Let $E_f$ and $E_b$ be functions defined by
\bea
E_f(x) &=& \frac{1}{e^x+1}=\frac{1}{2}\left[1-\tanh\left(\frac{x}{2}\right)\right] ,
%=\frac{e^{-x}}{1+e^{-x}},
\\[10pt]
E_b(x) &=& \frac{1}{e^x-1}=\frac{1}{2}\left[\coth\left(\frac{x}{2}\right)-1\right],
%\\[10pt]
%=\frac{e^{-x}}{1-e^{-x}}.
%E_0(x)&=&\frac{1}{2\sinh x}.
\eea
Notice that
\bea
%E_f(x)E_b(x) &=&E_b(2x),
%\\
E_f(x)&=& E_b(x)-2E_b(2x)\,.
%\\
%E_0(x)&=&\frac{1}{2}\left[E_f(x)+E_b(x)\right]=E_b(x)-E_b(2x)\,.
\eea

These functions can be represented as series which converge for any
$x>0$ 
\bea
E_f(x) &=& \sum_{k=1}^\infty (-1)^{k+1} e^{-kx},
\label{110xxc}
\\
E_b(x) &=& \sum_{k=1}^\infty e^{-kx},
\label{111xxc}
%\\
%E_0(x) &=& \sum_{k=0}^\infty \exp[-(2k+1)x].
%\label{110xxcz}
\eea
Now, by using eqs. (\ref{31bb})
%and (\ref{29xx})
we can rewrite these functions in the form
\bea
E_{b,f}[\beta(\omega-\mu)] &=& \int_0^\infty dt\; h_{b,f}(t,\beta\mu) 
\exp\left(-t\beta^2\omega^2\right),
\label{311xx}
\eea
where 
\bea
h_f(t,\beta\mu) &=& (4\pi)^{-1/2}t^{-3/2}
\sum_{k=1}^\infty (-1)^{k+1} k\exp\left(-\frac{k^2}{4t}+k\beta\mu\right),
\\
h_b(t,\beta\mu) &=& (4\pi)^{-1/2}t^{-3/2}
\sum_{k=1}^\infty k\exp\left(-\frac{k^2}{4t}+k\beta\mu\right),
\eea
It is easy to see that these functions satisfy the relations
\bea
h_f(t,\beta\mu)&=&h_b(t,\beta\mu)-\frac{1}{2}h_b\left(\frac{t}{4},\beta\mu\right).
\eea
To compute the asymptotics of the quantum heat traces we will need the asymptotics of the functions 
$h_{b,f}(t,\beta\mu)$ as $t\to 0$ and $t\to\infty$.
First of all, it is easy to see that as $t\to 0$
\bea
h_{b,f}(t,\beta\mu) \sim 
(4\pi)^{-1/2}t^{-3/2}
\exp\left(-\frac{1}{4t}+\beta\mu\right).
\eea
Therefore, the integral (\ref{314zz}) defining the quantum heat traces converges
at $t\to 0$.
The asymptotics of these functions as $t\to \infty$ (and $\mu<0$)
are computed as follows
\bea
h_f(t,\beta\mu) &\sim & (4\pi)^{-1/2}t^{-3/2}\frac{e^{\beta\mu}}{(e^{\beta\mu}+1)^2},
\label{42bb}
\\
h_b(t,\beta\mu) & \sim& (4\pi)^{-1/2}t^{-3/2}\frac{e^{\beta\mu}}{(e^{\beta\mu}-1)^2}.
%\\
%h_0(t,a)&\sim& (4\pi)^{-1/2}t^{-3/2}e^a\frac{e^{2a}+1}{(e^{2a}-1)^2}.
\eea
Therefore, the integral (\ref{314zz}) converges at $t\to\infty$ provided
the operator $H$ is positive.

The case $\mu=0$ is more complicated. For the fermionic case the eq. (\ref{42bb})
also holds for $\mu=0$; one just takes the limit $\mu\to 0^-$ to get
the asymptotics as $t\to \infty$,
\be
h_f(t,0) \sim  \frac{1}{8\sqrt{\pi}}t^{-3/2}.
\ee
In the bosonic case the limit $\mu\to 0$ is singular.
By carefully examining the 
behavior of the function $h_{b}$ we obtain as $t\to \infty$,
\be
h_b(t,0) \sim \frac{1}{\sqrt{\pi}}t^{-1/2}.
\ee
More generally, by using the Taylor expansions
\bea
\tanh x &=&\sum_{k=1}^\infty \frac{2^{2k}(2^{2k}-1)B_{2k}}{(2k)!}x^{2k-1},
\\
{}\coth x&=&\frac{1}{x}
+\sum_{k=1}^\infty \frac{2^{2k}B_{2k}}{(2k)!}x^{2k-1},
\eea
where $B_k$ are the Bernulli numbers, one can compute the asymptotic
expansion of the functions $h_{b,f}(t,0)$ as $t\to \infty$;
we obtain
\bea
h_{f}(t,0) &\sim&(4\pi)^{-1/2}
\sum_{k=1}^\infty (-1)^{k+1}\frac{(2^{2k}-1)B_{2k}}{2^{2k}k!}t^{-k-1/2},
\\[10pt]
h_{b}(t,0) &\sim& \frac{1}{\sqrt{\pi}}t^{-1/2}
-(4\pi)^{-1/2}\sum_{k=1}^\infty(-1)^{k+1}
\frac{B_{2k}}{2^{2k-1}k!}
t^{-k-1/2}.
\eea

The quantum heat traces 
%(from now on we set the chemical potential to zero, $\mu=0$)
now take the form
\bea
\Theta_{b,f}(\beta,\mu) &=&\Tr E_{b,f}[\beta(\omega-\mu)].
\label{34zz}
\eea
%We assume as usual that the mass parameter $m$ is sufficiently large 
%so that the  operator $\omega=(H_0+m^2)^{1/2}$ is positive.
In particular, there is a relation
\be
\Theta_f(\beta,\mu)=\Theta_b(\beta,\mu)-2\Theta_b(2\beta,\mu)\,.
\ee
Thus, by using eqs. (\ref{34zz}) and (\ref{311xx}) we obtain
for the quantum heat traces 
\bea
\Theta_{b,f}(\beta,\mu) &=& \int_0^\infty dt\; 
h_{b,f}\left(t,\beta\mu\right)
\Theta\left(t\beta^2\right),
\label{314zz}
\eea
which reduces the calculation of the quantum heat traces to the calculation of the
classical one (\ref{214zz}).
By using this representation we get for the quantum zeta functions
\bea
Z_{b,f}(s,\mu) &=&
\int_0^\infty dt\;
G_{b,f}(t,s,\mu)
\Theta\left(t\right),
\eea
where
\be
G_{b,f}(t,s,\mu)=\frac{1}{\Gamma(s)}\int_0^\infty d\beta \; \beta^{s-3}
h_{b,f}\left(\frac{t}{\beta^2},\beta\mu\right).
\label{316zz}
\ee

Another representation of the quantum heat traces can be obtained as follows.
Let $f(z)$ be a function of a complex variable $z$ such that it is analytic in
the region ${\rm Re}\,z>\gamma$, with some real parameter $\gamma$, and
decreases sufficiently fast at infinity in the right half-plane. Then the
function $f(-ip)$ of a complex variable $p$ is analytic in the region $\Imag
p>\gamma$ and decreases sufficiently fast at infinity in the upper half-plane.
Let $C$ be a contour in the upper half-plane of the complex variable $p$ that
goes from $-\infty+i\gamma$ to $\infty+i\gamma$ above all singularities of the
function $f(-ip)$. Then for any $x>0$ there holds
\be
f(x)=\frac{1}{2\pi i}\int_C
dp\;\frac{2p}{p^2+x^2}f(-ip);
\ee
with the only singularity above the contour $C$ being the simple pole at
$p=ix$. This can also be rewritten in the form
\be
f(x)=\int_0^\infty dt\; h(t)\exp(-tx^2),
\ee
where
\be
h(t)=\frac{1}{\pi i}\int_C
dp\;p f(-ip)\exp(-tp^2).
\ee

By using this equation we can represent the functions $E_{b,f}$ in the form
\be
E_{b,f}[\beta(\omega-\mu)]=\frac{1}{2\pi i}\int_C
dp\;\frac{2p}{p^2+\beta^2\omega^2}\;\frac{1}{\exp(-ip-\beta\mu)\mp 1}.
\label{325bb}
\ee
Here $C$ is a $\vee$ shaped contour going from $e^{i3\pi/4}\infty$ to $0$ and
then from $0$ to $e^{i\pi/4}\infty$ encircling only one pole at $i\beta\omega$. The
integrand has a sequence $\{p_k\}_{k\in\ZZ}$, of other simple poles:
\be
p_k=2k\pi+i\beta\mu,
\ee
in the bosonic case
and
\be
p_k=(2k+1)\pi+i\beta\mu
\ee
in the fermionic case. Therefore, for $\mu<0$ all other poles of the integrand
are in the lower half-plane and the contour of integration $C$ can be deformed
just to the real axis (with the integral understood in the {\it principal
value} sense).

Now, by using the integral representation (\ref{325bb})  
we get the same eq. (\ref{311xx})
with
\bea
h_{b,f}(t,\beta\mu) &=& \frac{1}{2\pi i}\int_C dp\; 
\frac{2p}{\exp(-ip-\beta\mu)\mp 1}\exp(-tp^2).
\eea
For $\mu<0$ this can be written as
\bea
h_{f}(t,\beta\mu) &=& \frac{1}{2\pi }\int_\RR dp\; p
\tan\left(\frac{p-i\beta\mu}{2}\right)\exp(-tp^2)
\nonumber\\
&=&\frac{1}{2\pi }\int_\RR dp\; 
\frac{p\sin p}{\cosh (\beta\mu)+\cos p}\exp(-tp^2),
\\
h_{b}(t,\beta\mu) &=& \frac{1}{2\pi }\int_\RR dp\;p
\cot\left(\frac{p-i\beta\mu}{2}\right)\exp(-tp^2)
\nonumber\\
&=&\frac{1}{2\pi }\int_\RR dp\; 
\frac{p\sin p}{\cosh(\beta\mu)-\cos p}\exp(-tp^2).
\eea
These equations also hold in the limiting case of $\mu\to 0$, with the
integrals taken in the principal value sense; the imaginary part cancels out.

Note that in the fermionic case none of the poles $p_k$ lie on the imaginary axis.
Therefore, for $\mu>0$ in the fermionic case the contour $C$ goes above all the
poles $p_k$ in the upper left half-plane, goes unterneath the pole at
$p=i\omega$, and then goes again above the poles $p_k$ in the right upper
half-plane. Another important observation is that
\be
\Real p_k^2=(2k+1)^2\pi^2-\beta^2\mu^2,
\ee
and, therefore, for sufficiently small $\beta\mu$, more precisely, for
$\beta\mu<\pi$, the real part of $p_k^2$ is positive for all $k$. Moreover, in
this case the contour $C$ can be deformed to a
${}^-\hspace{-5pt}\cup\hspace{-5pt}{}^-$ shaped contour which goes in the
region where $\Real p^2>0$ along the contour. For example, it can go from
$-\infty+i(\beta\mu+\varepsilon)$ to $-\pi+i(\beta\mu+\varepsilon)$, then from
$-\pi+i(\beta\mu+\varepsilon)$ to $0$, then from $0$ to
$\pi+i(\beta\mu+\varepsilon)$ and finally from $\pi+i(\beta\mu+\varepsilon)$ to
$\infty+i(\beta\mu+\varepsilon)$; with $\varepsilon$ an infinitesimal
parameter. For $\mu>0$ the function $h_f$ is given by the same formula
\bea
h_{f}(t,\beta\mu) &=& \frac{1}{2\pi }\int_C dp\; p\tan\left(\frac{p-i\beta\mu}{2}\right)\exp(-tp^2)
\eea
with a contour described above.

%=====================================
\section{Asymptotics of Quantum Heat Traces}
\setcounter{equation}0

Now, by substituting the heat kernel ansatz (\ref{421bb}) 
into the quantum heat traces
(\ref{314zz}) one can compute the integral over $t$ to obtain
the Mellin-Barnes representation of the quantum heat traces
\bea
\Theta_{b,f}(\beta,\mu) &=& 
2(4\pi)^{-(n+1)/2}
\frac{1}{2\pi i}\int\limits_{c-i\infty}^{c+i\infty}dq\;
\Gamma(-q)\Gamma\left[-q+(n+1)/2\right]\left(\frac{\beta}{2}\right)^{2q-n}
\nonumber\\
&&\times
F_{b,f}(n-2q,\beta\mu)A_q,
\label{597abb}
\eea
where $c<0$ and
\bea
F_b(s,{\beta\mu})&=&\sum_{k=1}^\infty\frac{e^{k{\beta\mu}}}{k^{s}},
\\
F_f(s,{\beta\mu})&=&\sum_{k=1}^\infty(-1)^{k+1}\frac{e^{k{\beta\mu}}}{k^{s}}.
\eea

For $\mu<0$ these series converge for any $s$ and define entire
functions of $s$.
For ${\rm Re}\,s>0$ they are given by the integrals
\bea
F_b(s,{\beta\mu})&=&\frac{1}{\Gamma(s)}\int_0^\infty dt\;\frac{t^{s-1}}{e^{t-{\beta\mu}}-1},
\\
F_f(s,{\beta\mu})&=&\frac{1}{\Gamma(s)}\int_0^\infty dt\;\frac{t^{s-1}}{e^{t-{\beta\mu}}+1},
\eea
Indeed, this can be proved by an expansion of the denominator
and computing the standard integral canceling the gamma function.
The analytic continuation for $\Real s>-N$ with any positive integer 
$N\in\ZZ_+$
is provided by the 
integration by parts
\bea
F_f(s,{\beta\mu})&=&\frac{1}{\Gamma(s+N)}\int_0^\infty dt\; t^{s+N-1}(-\partial_t)^N\frac{1}{e^{t-{\beta\mu}}+1},
\eea
and similarly for $F_b$.

Also, it is easy to see that
in the limit $\mu\to 0^-$ these functions are determined by the Riemann zeta function
\bea
F_b(s,0)&=&\zeta(s),
\label{421zz}
\\
F_f(s,0)&=&\left(1-2^{1-s}\right)\zeta(s).
\label{422zz}
\eea
Note that the fermionic function $F_f(s,0)$ is entire
but the bosonic function $F_b(s,0)$ is meromorhic
with a single pole at $s=1$.

For $\mu>0$ these series diverge for any $s$, however, they can be analytically 
continued 
to the polylogarithm,
 ${\rm Li}_s(z)$, \cite{gradshteyn14,erdelyi53}
 \bea
F_b(s,\beta\mu)={\rm Li}_{s}(e^{{\beta\mu}}),
\qquad
F_f(s,\beta\mu)=-{\rm Li}_{s}(-e^{{\beta\mu}}).
\eea
The integral for $F_b(s,\beta\mu)$ 
converges for $\Real s>0$ for any $\mu$ not lying on the positive real axis
and defines an analytic function of $\mu$ with infinitely many cuts 
along the horizontal lines $\beta\mu=x+2\pi i n$, with $x\ge 0$ and $n\in\ZZ$,
including the positive real axis. 
So, it has a jump across the positive real axis. Therefore,
$F_b(s,\beta\mu)$ is not well defined for real $\mu>0$.

The analytic structure of the function $F_f(s,\beta\mu)$ is similar.
It is an analytic function of $\mu$
with infinitely many cuts 
along the horizontal lines
$\beta\mu=x+\pi i (2n+1)$ with $x\ge 0$, $n\in\ZZ$.
 Therefore, the function $F_f(s,\beta\mu)$ is an analytic
function of $\mu$  in the neighborhood
of the real axis for $\Real s\ge 0$.
So, it is well defined for real positive $\mu$.
Therefore, for any real $\mu$ the fermionic function $F_f(s,\beta\mu)$
is an entire function of $s$.

For small $\mu$, to be precise for $\vert\beta\mu\vert<2\pi$, there
is an expansion \cite{gradshteyn14}: for non-integer $s\ne 1,2,3,\dots$,
\be
F_b(s,\beta\mu) = \Gamma(1-s)(-\beta\mu)^{s-1}
+\sum_{k=0}^\infty\frac{\zeta(s-k)}{k!}(\beta\mu)^{k},
\ee
and for positive integer $s=j=1,2,3,\dots$
\be
F_b(j,\beta\mu) = \left[\psi(j)+\CC-\log(-\beta\mu)\right]
\frac{(\beta\mu)^{j-1}}{(j-1)!}
+\sum_{k=0, k\ne j-1}^\infty\zeta(s-k)\frac{(\beta\mu)^{k}}{k!}.
\ee

%=========================================
% \mu<0
%

\subsection{Negative Chemical Potential}

The eq. (\ref{597a}) is especially useful for studying the asymptotic expansion
as $\beta\to 0$. We assume first that $\mu<0$. Then the integral
(\ref{597a}) is exactly a Mellin-Barnes integral studied in (\ref{52bb}), (\ref{53bb}).
The only difference with the relativistic heat trace (\ref{597a}) is the presence of the
functions $F_{b,f}$.
Therefore, it can be computed by the same method using Lemma 1. 
We decompose the heat traces in three parts
\be
\Theta_{b,f}(\beta,\mu)=\Theta^{\rm sing}_{b,f}(\beta,\mu)
+\Theta^{\rm loc}_{b,f}(\beta,\mu)+\Theta^{\rm non-loc}_{b,f}(\beta,\mu),
\label{612vvv}
\ee
and obtain
in even dimension $n=2m$,
\bea
\Theta^{\rm sing}_{b,f}(\beta,\mu)&\sim& 
\frac{1}{2}
(4\pi)^{-m+1/2}
\sum_{k=0}^m (-1)^k\frac{(2m-2k)!}{k!(m-k)!}\frac{1}{\beta^{2m-2k}}
F_{b,f}(2m-2k,\beta\mu)A_k,
\nonumber
\\
\\
\Theta^{\rm loc}_{b,f}(\beta,\mu)&\sim&
(-1)^m\frac{1}{2}(4\pi)^{-m}
\sum_{k=0}^\infty \frac{k!}{(k+m+1)!(2k+1)!}\beta^{2k+2}
\nonumber\\
&&\times
F_{b,f}(-2k-2,\beta\mu)A_{k+m+1},
\\[10pt]
\Theta^{\rm non-loc}_{b,f}(\beta,\mu)
&\sim&
-(-1)^m\pi^{-m}
\sum_{k=0}^\infty \frac{(k+m)!}{k!(2k+2m+1)!}\beta^{2k+1}
F_{b,f}(-2k-1,\beta\mu)A_{k+m+1/2},
\nonumber
\\
\eea
and in odd dimension $n=2m+1$,
\bea
\Theta^{\rm sing}_{b,f}(\beta,\mu)&\sim& 
2(4\pi)^{-m-1}
\sum_{k=0}^m (-1)^k\frac{(m-k)!}{k!}\left(\frac{2}{\beta}\right)^{2m-2k+1}
F_{b,f}(2m+1-2k,\beta\mu) A_k,
\nonumber
\\
\\
\Theta^{\rm loc}_{b,f}(\beta,\mu)&\sim&
(-1)^m2(4\pi)^{-m-1}
\sum_{k=0}^\infty \frac{1}{k!(k+m+1)!}
\left(\frac{\beta}{2}\right)^{2k+1}
\Biggl\{2F'_{b,f}(-2k-1,\beta\mu)
\nonumber\\%[10pt]
&&\hspace{-1cm}
+\left[\psi(k+m+2)+\psi(k+1)-2\log(\beta/2)\right]
%\nonumber\\
%&&\times
F_{b,f}(-2k-1,\beta\mu)
\Biggr\}A_{k+m+1},
\nonumber\\
\\
\Theta^{\rm non-loc}_{b,f}(\beta,\mu)&\sim& 
-(-1)^m2(4\pi)^{-m-1}
\sum_{k=0}^\infty \frac{1}{k!(k+m+1)!}\left(\frac{\beta}{2}\right)^{2k+1}
\nonumber\\
&&\times
F_{b,f}(-2k-1,\beta\mu)A'_{k+m+1}.
%\nonumber\\
\eea
It is worth mentioning that since the fermionic function $F_f(s,\beta\mu)$
is entire for any real $\mu$ the formulas for the fermionic quantum heat trace
are valid also for zero and positive chemical potential $\mu$.

%============================
%=====================================
% \mu=0
%

\subsection{Zero Chemical Potential}

Now, we study the limit of zero chemical potential $\mu\to 0^-$. In this case
the functions $F_{b,f}(s,\beta\mu)$ are given by  
(\ref{421zz}) and (\ref{422zz}). So, we obtain
\bea
\Theta_{b}(\beta,\mu) &=& 
2(4\pi)^{-(n+1)/2}
\frac{1}{2\pi i}\int\limits_{C_0}dq\;
\Gamma(-q)\Gamma\left[-q+(n+1)/2\right]\left(\frac{\beta}{2}\right)^{2q-n}
\nonumber\\
&&\times
\zeta(n-2q)A_q,
\label{597abbzz}
\eea
and
\bea
\Theta_{f}(\beta,\mu) &=& 
2(4\pi)^{-(n+1)/2}
\frac{1}{2\pi i}\int\limits_{C_0}dq\;
\Gamma(-q)\Gamma\left[-q+(n+1)/2\right]\left(\frac{\beta}{2}\right)^{2q-n}
\nonumber\\
&&\times
(1-2^{1-n+2q})\zeta(n-2q)A_q,
\label{597abbz}
\eea
Of course, we could have obtained the same results from eqs.
(\ref{328bbb})-(\ref{329bbb}) and (\ref{47bbb}).

%=======================================================

First of all, we recall that the fermionic function $F_f(s,0)$
is entire. Therefore, there are no additional singularities in the integral
and we can just substitute $\mu=0$ in the formulas
of the previous section. 
We use the same decomposition of the quantum heat traces (\ref{612vvv})
and the fact that the
values of the Riemann zeta function at negative even integers
vanish \cite{erdelyi53}
\be
\zeta(-2k) = 0, \qquad
k=1,2,3,\dots
\ee
Therefore, the function $\Theta^{\rm loc}_f(\beta,0)$ 
does not contribute to the asymptotic expansion in even dimension
$n=2m$
\be
\Theta^{\rm loc}_f(\beta,0) \sim 0.
\ee
For the other parts we obtain
in even dimension $n=2m$,
\bea
\Theta^{\rm sing}_{f}(\beta,0)&\sim& 
\frac{1}{2}
(4\pi)^{-m+1/2}
\sum_{k=0}^m (-1)^k\frac{(2m-2k)!}{k!(m-k)!}\frac{1}{\beta^{2m-2k}}
\nonumber
\\
&&\times\left(1-2^{2k-2m+1}\right)\zeta(2m-2k)A_k,
\\
\Theta^{\rm non-loc}_{f}(\beta,0)
&\sim&
-(-1)^m\pi^{-m}
\sum_{k=0}^\infty \frac{(k+m)!}{k!(2k+2m+1)!}\beta^{2k+1}
\nonumber
\\
&&\times\left(1-2^{2k+2}\right)\zeta(-2k-1)
A_{k+m+1/2},
\eea
and in odd dimension $n=2m+1$,
\bea
\Theta^{\rm sing}_{f}(\beta,0)&\sim& 
2(4\pi)^{-m-1}
\sum_{k=0}^m (-1)^k\frac{(m-k)!}{k!}\left(\frac{2}{\beta}\right)^{2m-2k+1}
\nonumber\\
&&\times\left(1-2^{2k-2m}\right)\zeta(2m-2k+1)A_k,
%\nonumber
%\\
\\
\Theta^{\rm loc}_{f}(\beta,0)&\sim&
(-1)^m2(4\pi)^{-m-1}
\sum_{k=0}^\infty \frac{1}{k!(k+m+1)!}
\left(\frac{\beta}{2}\right)^{2k+1}
\nonumber\\%[10pt]
&&
\hspace{-1cm}
\times\Biggl\{\left[\psi(k+m+2)+\psi(k+1)-2\log(\beta/2)\right]
%\nonumber\\
%&&\times
\left(1-2^{2k+2}\right)\zeta(-2k-1)
\nonumber\\
&&
\hspace{-1cm}
+2(1-2^{2k})\zeta'(-2k-1)
-2^{2k+1}\zeta(-2k-1)\log 2
\Biggr\}A_{k+m+1},
%\nonumber\\
\\
\Theta^{\rm non-loc}_{f}(\beta,0)&\sim& 
-(-1)^m2(4\pi)^{-m-1}
\sum_{k=0}^\infty \frac{\left(\beta/2\right)^{2k+1}}{k!(k+m+1)!}
\nonumber\\
&&\times
\left(1-2^{2k+2}\right)\zeta(-2k-1)A'_{k+m+1}.
\nonumber\\
\eea

%===============================

All we need to do now is to compute
the values of the zeta function and its
derivatives at negative integer points. It is well known that
the values of the zeta function at negative integers
are determined by the Bernoulli numbers
\cite{erdelyi53}
\bea
\zeta(-2k+1)&=&
(-1)^k\frac{2(2k-1)!}{(2\pi)^{2k}}\zeta(2k)
=-\frac{B_{2k}}{2k}.
\eea
Also, by using the functional equation for the zeta function we obtain
the derivatives at negative integer points
\bea
\zeta'(-2k)&=&(-1)^k\frac{(2k)!}{2(2\pi)^{2k}}\zeta(2k+1)
\\
\zeta'(-2k+1)&=&-(-1)^{k}\frac{2(2k-1)!}{(2\pi)^{2k}}
\left\{\zeta'(2k)+\left[\psi(2k)-\log(2\pi)\right]\zeta(2k)\right\}
\eea

%======================================================

Now we turn to the bosonic case. In this case the function
$F_b(s,0)= \zeta(s)$ acquires an additional pole at $s=1$. This leads
to an additional pole of the integrand at 
$q=(n-1)/2$ which needs
to be taken into account when computing the contour integral
(\ref{597abbzz}).
In even dimensions this pole is simple, however, in odd dimensions this pole
coincides with one of the poles of one of the gamma functions, making it a
double pole. 

Therefore our Lemma 1 needs a modification.
We consider the cases of even and odd dimensions
separately.
In even dimension $n=2m$ w
we still have
\be
\Theta^{\rm loc}_b(\beta,0) \sim 0
\ee
and we just need to compute an extra term due to
an additional simple pole. 
The extra term is easily computed by the residue calculus.
We obtain
\be
\Theta_{b}(\beta,0)=\Theta^{\rm sing}_{b}(\beta,0)
+S(\beta)
+\Theta^{\rm non-loc}_{b}(\beta,0),
\label{612wwwz}
\ee
where
\bea
S(\beta)&=&
(-1)^m\pi^{-m}\frac{m!}{(2m)!}\frac{1}{\beta}A_{m-1/2}
\eea
is the additional term. The other terms are:
\bea
\Theta^{\rm sing}_{b}(\beta,0)&\sim& 
\frac{1}{2}
(4\pi)^{-m+1/2}
\sum_{k=0}^m (-1)^k\frac{(2m-2k)!}{k!(m-k)!}\frac{1}{\beta^{2m-2k}}
\nonumber
\\
&&\times\zeta(2m-2k)A_k,
\\
\Theta^{\rm non-loc}_{b}(\beta,0)
&\sim&
-(-1)^m\pi^{-m}
\sum_{k=0}^\infty \frac{(k+m)!}{k!(2k+2m+1)!}\beta^{2k+1}
\nonumber
\\
&&\times\zeta(-2k-1)
A_{k+m+1/2}.
\eea

In odd dimension $n=2m+1$ the
zeta function has a pole at $q=m$ which coincides with a pole
of one of the gamma functions making it a double pole.
So, we need to remove it from
the previous sum and add the recomputed correct term.
This will be the terms proportional to $A_m$ and $A'_m$.
We obtain
\be
\Theta_{b}(\beta,0)=\tilde\Theta^{\rm sing}_{b}(\beta,0)
+\tilde S(\beta)
+\Theta^{\rm loc}_{b}(\beta,0)
+\Theta^{\rm non-loc}_{b}(\beta,0),
\label{612www}
\ee
where
\bea
\tilde\Theta^{\rm sing}_{b}(\beta,0)&\sim& 
2(4\pi)^{-m-1}
\sum_{k=0}^{m-1} (-1)^k\frac{(m-k)!}{k!}\left(\frac{2}{\beta}\right)^{2m-2k+1}
\zeta(2m+1-2k) A_k,
\nonumber
\\
\label{637zz}
\\
\Theta^{\rm loc}_{b}(\beta,0)&\sim&
(-1)^m2(4\pi)^{-m-1}
\sum_{k=0}^\infty \frac{1}{k!(k+m+1)!}
\left(\frac{\beta}{2}\right)^{2k+1}
\nonumber\\
&&\times
\Biggl\{
\left[\psi(k+m+2)+\psi(k+1)-2\log(\beta/2)\right]\zeta(-2k-1)
\nonumber\\
&&
%\times
+2\zeta'(-2k-1)
\Biggr\}A_{k+m+1},
%\nonumber\\
\\
\Theta^{\rm non-loc}_{b}(\beta,0)&\sim& 
-(-1)^m2(4\pi)^{-m-1}
\sum_{k=0}^\infty \frac{1}{k!(k+m+1)!}\left(\frac{\beta}{2}\right)^{2k+1}
%\nonumber\\
%&&\times
\zeta(-2k-1)A'_{k+m+1}.
\nonumber\\
\eea
Here we removed the last term from the sum for $k=m$
in eq. (\ref{637zz}) since it would be singular as $\zeta(1)$
and recomputed it. The correct term is
\be
\tilde S(\beta) =
-2\frac{(-1)^m}{m!}(4\pi)^{-m-1}\frac{1}{\beta}
\Biggl\{A'_m-\left[\psi(m+1)-\psi(1)\right]A_m
+2\log(\beta/2)A_m\Biggr\}
%\nonumber
\ee

Note that the extra term is proportional to
$\beta^{-1}$ and $\beta^{-1}\log\beta$ and is
determined by the invariant $A_{(n-1)/2}$ and
its derivative $A'_{(n-1)/2}$ at the point $q=(n-1)/2$.
This invariant is like the zeta regularized determinant
(which is proportional to $A_{n/2}$ and 
$A'_{(n-1)/2}$); it is global and cannot be computed 
locally in a generic case. For the lack of a better name we call this new 
invariant the {\it residue}.

%=================================================================
\section{Conclusion}
\setcounter{equation}0

The primary goal of this paper was to introduce and to study
some new invariants of second-order
elliptic partial differential operators
on manifolds, $\Theta_{b,f,r}(\beta,\mu)$ given by
(\ref{214zzbb}) and (\ref{219vv})-(\ref{220vv})
that we call the relativistic heat trace
and the quantum heat traces.
Of special interest are the asymptotics of these invariants 
as $\beta\to 0$.
First, we showed how these heat traces
can be reduced to the some integrals of the standard heat trace.
Then, by using a special Mellin-Barnes representation of the
heat kernel for the Laplace type operator 
introduced
in our paper \cite{avramidi91}
we were able to compute
the asymptotics of the quantum heat traces.
We showed that the asymptotic expansion has both power and
logarithmic terms.
We expressed the
coefficients of the asymptotic expansion in terms of values of 
an entire function $A_q$ and its derivatives at integer and half-integer
points. The values of this entire function at the positive integer
points $A_k$ are equal to the standard heat trace coefficients and are
locally computable, while the values of the function at half-integer
points, $A_{k+1/2}$, as well as the derivatives of it at integer points,
$A'_k$,
are non-trivial
global invariants.

\section*{Acknowledgments}
\setcounter{equation}0

The financial support 
by the Alexander von Humboldt Foundation and the Federal Ministry for 
Education and Research of Germany is greatfully acknowledged.
I would also like to thank the Mathematical Institute of the University of Bonn, in particular, Werner M\"uller
and other colleagues for hospitality. 

%======================================================

%===================================================
\end{document}